\documentclass[twocolumn,floatfix,showpacs,preprintnumbers,amsmath,amssymb,prb,superscriptaddress]{revtex4}

\usepackage{graphicx}
\usepackage{dcolumn}
\usepackage{bm}

\begin{document}

\title{Self-assembly of monodisperse clusters: Dependence on target geometry}

\author{Alex W.~Wilber}
\affiliation{Physical and Theoretical Chemistry Laboratory,
  Department of Chemistry, University of Oxford, South Parks Road,
  Oxford, OX1 3QZ, United Kingdom}
\author{Jonathan P.~K.~Doye}
\thanks{Author for correspondence}
\affiliation{Physical and Theoretical Chemistry Laboratory,
  Department of Chemistry, University of Oxford, South Parks Road,
  Oxford, OX1 3QZ, United Kingdom}
\author{Ard A.~Louis}
\affiliation{Rudolf Peierls Centre for Theoretical Physics,
 University of Oxford, 1 Keble Road, Oxford, OX1 3NP, United Kingdom}

\date{\today}

\begin{abstract}
We apply a simple model system of patchy particles to study monodisperse 
self-assembly, using the Platonic solids as target structures. 
We find marked differences between the assembly behaviours of the different 
systems. 
Tetrahedra, octahedra and icosahedra assemble
easily, while cubes are more challenging and dodecahedra do not assemble. 
We relate these differences to the kinetics and thermodynamics of assembly,
with the formation of large disordered aggregates a particular important 
competitor to correct assembly.
In particular, the free energy landscapes of those targets that are 
easy to assemble
are funnel-like, whereas for the dodecahedral system the landscape is relatively
flat with little driving force to facilitate escape from disordered aggregates.
\end{abstract}

\pacs{81.16.Dn,47.57.-s,87.15.ak}

\maketitle

\section{Introduction}

The assembly of nano-structured materials and devices presents a great challenge for the future. A very promising approach is offered by self-assembly
processes, in which nanoscopic or colloidal building blocks come together spontaneously to form ordered structures. While top-down approaches to
assembly become increasingly challenging on small length scales, self-assembly offers a bottom-up approach which circumvents many of these difficulties,
while providing its own unique challenges in the synthesis and design of the subunits.

Current examples of synthetic self-assembling systems, such as micelles and block copolymers, usually lead to polydisperse and imperfectly controlled products.
Biological systems, by contrast, display an astonishing variety of ordered and precise self-assembly processes.\cite{Goodsell04} For example, virus
replication involves the assembly of hundreds of proteins to form highly symmetric and monodisperse shells (capsids).
Such systems provide an inspiring example of the level of control possible in self-assembly. Future applications in nanotechnology
are likely to require this level of sophisticated control in order to form precisely ordered structures.

One attractive aspect of self-assembly is the idea that the structure of the final product is entirely determined by the interactions between the subunits
(along with the reaction conditions). Hence, given sufficient understanding of 
the principles of self-assembly, the subunits can be 
designed to produce a given product. In order to implement these designs a great deal of control is required over the structure of the subunits. One key
requirement for sophisticated self-assembly is the production of anisotropic subunits, which only attract each other in certain directions.
Rational design of this anisotropy, both in the shape and in the interactions, 
then allows for control over the resultant self-assembled structures.

Recently there has been a great deal of progress towards synthesising such 
anisotropic subunits.\cite{Edwards07,Glotzer07b,Yang08} Promising avenues 
included nanoparticles covered with mixed thiol monolayers,\cite{Jackson04,Stellacci07}
two-faced Janus\cite{Roh05,Hong08} and even triphasic \cite{Roh06} particles 
created at interfaces,
small colloidal clusters\cite{Pine2003,Perro08} where the gaps between 
the individual colloids can be filled with a controlled amount of another material to create symmetrical `patchy' particles\cite{Cho05_Bidisperse,Cho07,Kraft09}
and colloids with patches grown at the sites of contacts in a crystal.\cite{Wang08}

The production of functional self-assembling systems will require not only advanced synthesis techniques to produce the subunits, but also a
strong theoretical understanding of the self-assembly process so that appropriate designs and conditions can be chosen. Recently there has been significant
progress in the theory\cite{Endres02,Endres05,Zlotnick05_TheoreticalAspects,Zandi06} and simulation\cite{Hagan06,Hagan07,Hagan08,Hagan08_2,Brooks07,Nguyen08,Nguyen08b,Schwartz06,Schwartz07,Schwartz08,Rapaport04,Rapaport08,Villar09,Zhang04,vanWorkum06,Wilber07,Glotzer07}
of monodisperse self-assembly.
However, these studies have mostly focussed on the self-assembly of virus capsids or other protein complexes,\cite{Endres02,Endres05,Zlotnick05_TheoreticalAspects,Hagan06,Hagan07,Hagan08,Hagan08_2,Brooks07,Nguyen08,Nguyen08b,Schwartz06,Schwartz07,Schwartz08,Rapaport04,Rapaport08,Villar09} and so include protein-like interactions which 
are more specific than those which are likely to be available in the first 
generation of synthetic patchy particles.

Here we consider a minimal patchy particle model, representing spherical colloids or nanoparticles with anisotropic interactions. Our model is intentionally simple,
so that we are easily able to observe and study behaviours
which we expect to be general properties of monodisperse self-assembly rather than being specific to our model.
This simplicity also allows us to comprehensively survey the behaviour, and to 
connect the kinetics with the form of the underlying free energy landscapes.

We have previously used the model to study the self-assembly of 12-particle 
icosahedra.\cite{Wilber07}  Here, we build on this work to consider the 
effect of the target geometry on the self-assembly process, using the Platonic solids
as our targets. We use these examples to address questions such as
whether self-assembly becomes more difficult with increasing target size, 
and whether particular geometric features make some targets particularly difficult 
to assemble. In doing so, we will pay particular attention to the main processes 
which compete with successful assembly, and how their impact on the yield can be 
minimised. We hope that the design principles we learn from this work will offer 
some guidance to experimental groups seeking to make practical synthetic 
self-assembling systems.

Although the assembly process studied here is similar to that of virus capsids, in that anisotropic particles come together to form closed,
highly symmetric shell structures, the interaction potential we use has no 
dependence on the torsional angle between interacting particles. 
While this is likely to be a good choice for modelling synthetic anisotropic particles, 
it is not a good representation of the interactions between proteins, 
and leads to behaviour not observed for systems of virus capsomers. We consider a model including torsional interactions, and which hence more closely mirrors 
capsid assembly, in the accompanying paper.\cite{accompanying}

\section{Methods}

\subsection{Model}
\label{sec:Model}

We make use of a minimal model, designed to contain only the essential features required for targeted self-assembly, while allowing for efficient
simulation. The model consists of spherical particles patterned with attractive patches. They are described by a modified Lennard-Jones potential,
in which the repulsive part of the potential is isotropic, but the attractive part is anisotropic and depends on the alignment of patches
on interacting particles. Specifically, the potential is described by
\begin{equation}
\label{eq:potential}
V_{ij}({\mathbf r_{ij}},{\mathbf \Omega_i},{\mathbf \Omega_j})=\left\{
    \begin{array}{ll}
       V_{\rm LJ}(r_{ij}) & r<\sigma_{\rm LJ} \\
       V_{\rm LJ}(r_{ij})
       V_{\rm ang}({\mathbf {\hat r}_{ij}},{\mathbf \Omega_i},{\mathbf \Omega_j})
                       & r\ge \sigma_{\rm LJ}, \end{array} \right.
\end{equation}
where $V_{\rm LJ}$, the Lennard-Jones potential, is given by
\begin{equation}\label{eqn:LJ} V_{\rm LJ}(r) = 4\epsilon\left[ \left( \frac{\sigma_{\rm LJ}}{r}
    \right)^{12} - \left( \frac{\sigma_{\rm LJ}}{r} \right)^{6} \right].
\end{equation}
$V_{\rm ang}$ is an angular modulation factor, which depends on the orientations of the patches on the two interacting particles, as well as
the direction of the vector joining them. Specifically,
\begin{equation}
V_{\rm ang}({\mathbf {\hat r}_{ij}},{\mathbf \Omega_i},{\mathbf \Omega_j})=
G_{ij}({\mathbf {\hat r}_{ij}},{\mathbf \Omega_i})
G_{ji}({\mathbf {\hat r}_{ji}},{\mathbf \Omega_j}),
\end{equation}
where
\begin{equation}
\label{eqn:AngMod}
G_{ij}({\mathbf {\hat r}_{ij}},{\mathbf \Omega_i})=
\exp\left(-\frac{\theta_{k_{\rm min}ij}^2}{2\sigma^2}\right),
\end{equation}
$\sigma$ gives the width of the Gaussian, $\theta_{kij}$ is the angle between patch vector $k$ on particle $i$
and the interparticle vector $\mathbf r_{ij}$, and $k_{\rm min}$ is the patch that minimizes the magnitude of this angle. Hence,
only the patches on each particle that are closest to the interparticle axis interact with each other, and $V_{\rm ang}$ is one if the patches point
directly at each other.
One feature of this potential is that as $\sigma\rightarrow\infty$ the isotropic Lennard-Jones potential is recovered.
For computational efficiency the potential is truncated and shifted at $r=3\,\sigma_{\rm LJ}$, and the crossover distance in Eq.\ \ref{eq:potential} 
is adjusted so that it still occurs when the potential is zero.

\begin{figure}
\includegraphics[width=8.4cm]{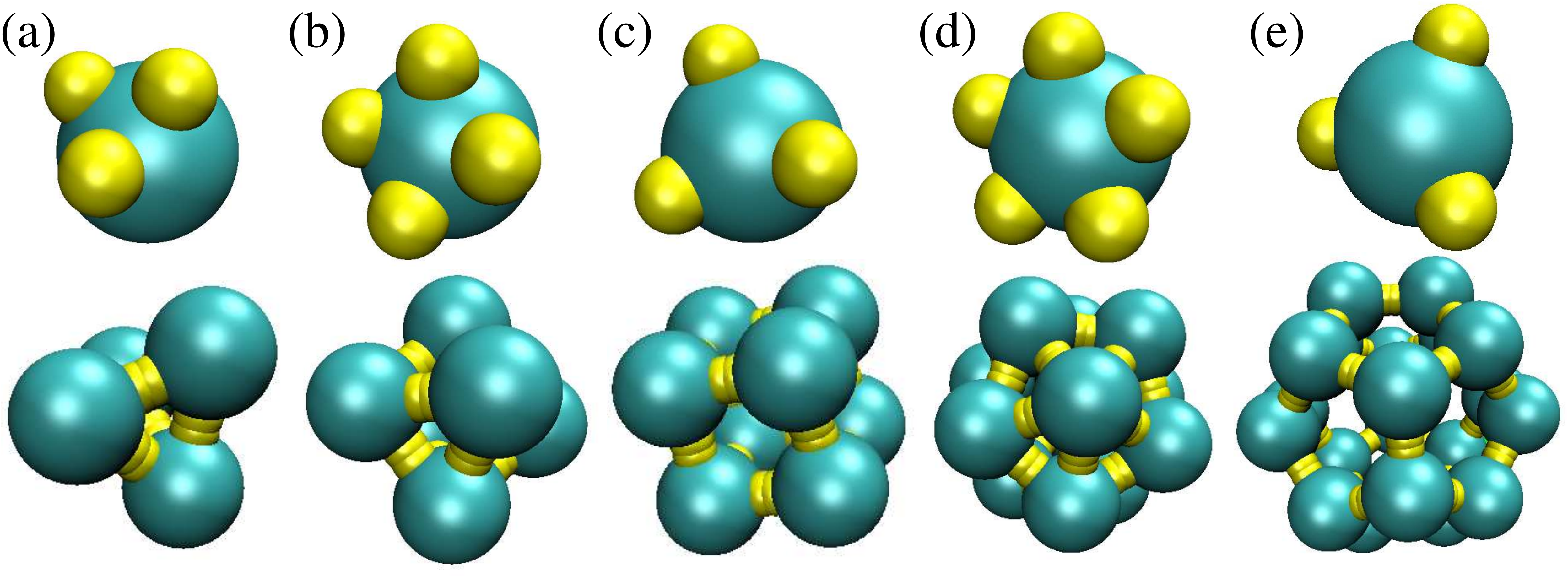}
\caption{\label{fig:ShapePics}(Colour Online) Single particles and complete clusters for the different target structures: (a) tetrahedron, (b) octahedron, 
(c) cube, (d) icosahedron and (e) dodecahedron.  
}
\end{figure}

A particular particle is specified by a set of unit vectors describing 
the positions of the attractive patches.
For each of our target structures, the patches are placed such that they
point directly at the neighbouring particles in the target structure. Fig.\ \ref{fig:ShapePics} shows the component particles and complete
clusters for each of our target structures, the Platonic solids. Note that 
for these targets all the particles and patches are equivalent.

Somewhat similar patchy particle models have been used to study gel formation,\cite{Bianchi06,Sciortino08} the crystallization of proteins\cite{Sear99c,Kern03} and patchy colloids\cite{Zhang06,Doye07} and fibre formation.\cite{Shiryayev06,Huisman08}

\subsection{Dynamical simulations}

In the simulations of our model we wish to represent the Brownian motion that
colloids and nanoparticles undergo in solution. As we do not include any solvent
particles in our coarse-grained description of the system, a simple and 
efficient way to represent this dynamics is to use Monte Carlo (MC) where the 
moves are restricted to be local, since this ensures that the dynamics are
diffusive.

In particular, we use Metropolis MC in the canonical ensemble, 
using periodic boundary conditions.
The allowed move types are small single-particle translations and rotations. The translational moves are randomly chosen from a cube centred
on the selected particle. Rotational moves make use of a quaternion description of the particle's orientation; the proposed quaternion
is given by the renormalized sum of the current quaternion and a smaller, randomly generated 4-vector.

One potential problem with using single-particle moves is that, although free
particles and clusters undergo diffusion as required, the relative diffusion
rates of clusters of different sizes can be incorrect with the larger clusters
diffusing too slowly. However, in practice for systems where the main mechanism
of cluster growth is by monomer addition rather than cluster-cluster 
aggregation, single-particle moves are sufficient. Indeed, preliminary 
simulations using the virtual move MC algorithm, which has been very recently 
introduced by Whitelam and Geissler and is designed to overcome this
problem by using cluster moves,\cite{Whitelam07,Whitelam08} 
show only very minor differences to those presented here. 
By contrast, we have found that such an algorithm is crucial
for systems designed to assemble hierarchically.

\subsection{Equilibrium simulations}
\label{sec:EquilSims}
To compute free energy landscapes and the positions of thermodynamic 
transitions 
we make use of umbrella sampling. The essential idea is to bias the system such that free energy barriers
are easier to cross, thus allowing the system to explore configuration 
space much more quickly. Under
this scheme, instead of choosing the acceptance probabilities using a Boltzmann distribution, we use the modified distribution
$\exp\left[-\beta\left(\mathcal{V}+w(Q)\right)\right]$, where $Q$ is an order parameter
and $w(Q)$ is a weighting function. Canonical averages are simple to obtain from a simulation of such a non-Boltzmann (nB) ensemble using the expression
\begin{equation} \left<B\right>_{NVT} = \left<B \exp \left[ \beta w(Q) \right] \right>_{\mathrm{nB}}, \end{equation}
where $B$ is some generic property of the system.

Despite considerable effort, it did not prove feasible to devise an order 
parameter that facilitated the formation of multiple copies of the target
structure. Therefore, we instead restricted our umbrella sampling 
simulations to the formation of a single target structure at the same density. 
There are, however, two possible reasons why the equilibrium data for this 
finite-sized system can differ from that for bulk. 
Firstly, it does not allow for interactions between the 
target clusters. But since the attractions between these clusters are 
small, this is a good approximation. Secondly, it does not allow for 
configurations where the fraction of the atoms in the target structure is 
non-integer; in our umbrella sampling simulations the only possibilities are 
for all the particles to be in the complete target structure or none of them. 
For example, at the midpoint of the transition, the small system will fluctuate in time between the target cluster and the monomeric gas state, spending half the time in each state, whereas for a large system one would expect that at any given time the configuration will be a mixture of target clusters and monomers with half the particles in the target clusters.
Analytical calculations have indicated that this restriction
has only a relatively small effect on the position of the centre of the 
transition between monomers and the target structure (this is what we 
are most interested in here), but can cause an appreciable narrowing of
the transition compared to the thermodynamic limit.\cite{TomOunpub}

A convenient order parameter to study the transition between a monomeric gas and the target structure is then the number of particles in the largest 
cluster in the simulation. 
The weighting function $w(Q)$ was found by iteratively performing simulations 
and improving $w(Q)$ at each iteration until the time spent
at each value of $Q$ was approximately equal.

In the case of the dodecahedron we found that the above order parameter was 
inadequate because the larger clusters that formed were disordered rather than
based on the dodecahedron. We therefore introduced a second-order parameter, 
namely the number of pentagons in the largest cluster, with the aim of aim
of promoting the formation of dodecahedral clusters.
However, this 
was still insufficient to drive the system to assemble successfully.

Instead, to finally achieve equilibrium we had to combine this
two-dimensional umbrella-sampling scheme with Hamiltonian exchange.\cite{Sugita00} 
In the latter, 
additional exchange moves are introduced that involve swapping configurations 
between simulations with different Hamiltonians, the purpose being to couple 
the system of interest to one where equilibrium is easier to achieve. 
In our case, we know that successful assembly of dodecahedra is possible for a 
similar patchy particle model, but where a torsional component is included 
in the potential.\cite{accompanying}
Specifically, the attractive region of the potential is modulated by an 
additional factor
\begin{equation}
V_{\rm tor} = \exp\left(-\frac{\phi}{2\sigma_{\rm tor}^2}\right),
\end{equation}
where $\phi$ is a torsional angle between two particles, chosen such that 
$\phi=0$ in the target structure.\cite{accompanying}
In the limit that $\sigma_{\rm tor}$ approaches infinity, our original 
potential  is recovered.

In our Hamiltonian exchange scheme we ran an array of 20 simulations at 
different values of $\sigma_{\rm tor}$ ranging from 1.25 radians to $\infty$. 
The intermediate values were chosen to maximise the rate of 
configurational exchange.
The probability of an exchange move was $0.1$ and only
exchanges between adjacent values of $\sigma_{\rm tor}$ were attempted.
The acceptance probability that ensures detailed balance is
\begin{equation}
\begin{split}
p_{\mathrm{acc}}((i,1),(j,2)\rightarrow(i,2),(j,1)) =  
\min\left\{1,\right. \\
\left. \exp\left[\beta(W_{i,1} + W_{j,2} - W_{i,2} - W_{j,1})\right]\right\}
\end{split}
\end{equation}
where $i$ and $j$ represent different Hamiltonians, 
and $1$ and $2$ different configurations. 
$W_{i,1}$ is the combined potential energy and umbrella weighting of
configuration $1$ with the Hamiltonian $i$, 
i.e.\ $W_{i,1} = V_i(\mathbf{r_1}^N) + w_i(Q(\mathbf{r_1}^N))$.

\subsection{Structural Analysis}
\label{sec:BondAndRingCounting}

Many of our results make use of information on the clusters present in the simulations, and of the rings present within those clusters. We define
two particles as being bonded if their interaction energy satisfies $V_{ij} \le -0.4$. Particles are part of the same cluster if they are connected
by an unbroken chain of bonds. We identify target clusters by matching the number of particles and bonds in the cluster to the profile for the
target cluster, including some allowance for bonds temporarily weakened by thermal fluctuation.
We define rings using the shortest-path algorithm given in Ref.\ 
\onlinecite{Yuan02}; a closed loop is counted as a ring
if there is no pair of particles on the ring connected by any chain of bonds shorter (i.e.\ consisting of fewer bonds) than the route by which
they are connected along the ring.

\section{Results}

\begin{figure}
\includegraphics[width=8.0cm]{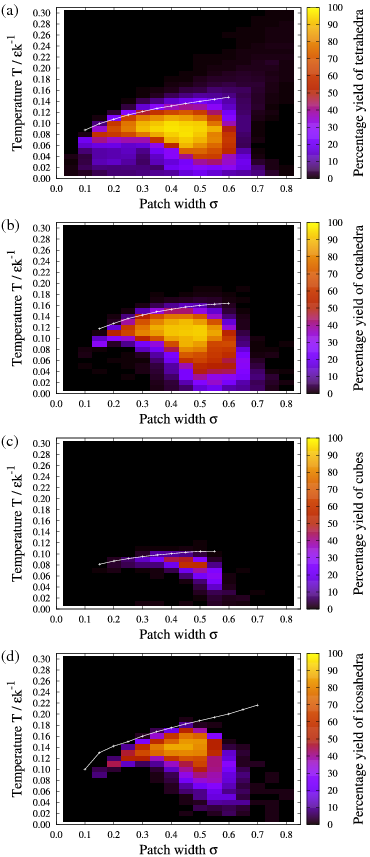}
\caption{\label{fig:MainHeatPlots}(Colour Online) The percentage yield of target clusters formed after $80\,000$ MC cycles
as a function of the patch width $\sigma$ (measured in radians) and the temperature for 1200 particles at a number density of
$0.15\,\sigma_{\rm LJ}^{-3}$, for particles designed to form 
(a) tetrahedra, (b) octahedra, (c) cubes and (d) icosahedra.
No plot is included for particles designed to form dodecahedra, 
since under no conditions did any dodecahedra ever assemble. 
The white lines show the thermodynamic transition temperature $T_c$ for the 
transition from a gas of clusters to a gas of monomers calculated using 
umbrella sampling.
}
\end{figure}

We first wish to identify the conditions under which successful assembly of
the target structure occurs. Because of the relative simplicity of our model, 
we are able to perform large numbers of simulations, and
hence comprehensively map out the assembly behaviour over a range of parameter 
space.  Fig.\ \ref{fig:MainHeatPlots} shows the dependence of the final
yields of each of the target structures on temperature (relative to $\epsilon$) and 
patch width in simulations starting from an initial random geometry after
a given number of MC cycles. Although the yields in certain regions will tend
to increase slightly with longer simulation times, the form of the diagrams
is robust. Also note that 
low (high) temperature is equivalent to strong (weak) interactions.

To first order the general shapes of the plots are similar.
because the same physical principles apply in each case.
However, there are also significant differences. 
Tetrahedra, octahedra and icosahedra assemble readily over wide ranges of parameter space. By contrast, cubes are more difficult to assemble and are only
formed in a smaller region of parameter space, while dodecahedra never form at all. We shall examine the reasons for these differences in detail later.

The region of successful assembly is determined by a number of constraints, both thermodynamic and kinetic. 
Firstly, we find that at high
temperature $T$ the stable state becomes a gas of monomers and small clusters, and target clusters are no longer formed. 
A second thermodynamic constraint arises at high patch width $\sigma$, 
where the patches are so wide that the target clusters cease to be the most 
stable state of the system. Each patch then becomes capable of interacting with 
more than one neighbour, and so the system can both lower its
energy and raise its entropy by forming large, unstructured liquid-like droplets.

\begin{figure}
\includegraphics[width=7.4cm]{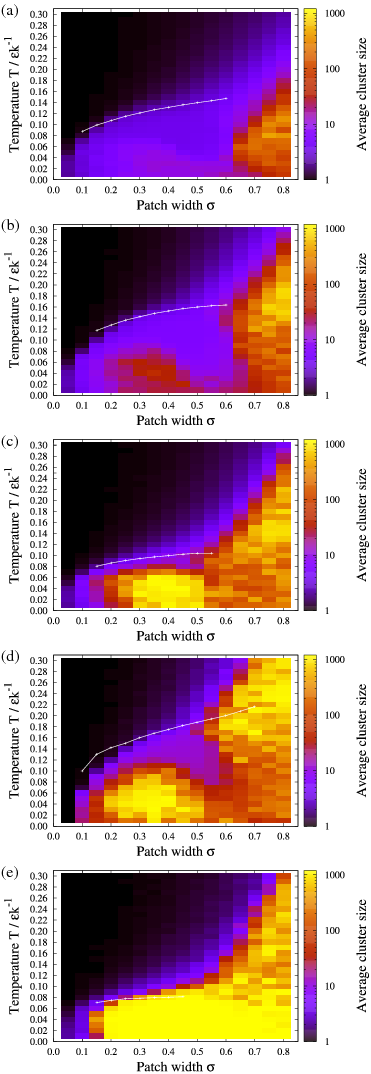}
\caption{\label{fig:SizeHeatPlots}(Colour Online) The mean cluster size (averaged over particles) of systems
of particles designed to form (a) tetrahedra, (b) octahedra, (c) cubes, (d) icosahedra, and (e) dodecahedra, for the same
simulations as Fig.\ \ref{fig:MainHeatPlots}. The white lines show 
$T_c$. 
}
\end{figure}

Both of these effects can be seen in Fig.\ \ref{fig:SizeHeatPlots}, which shows the average cluster size for each target structure,
again as a function of $T$ and $\sigma$. At high $T$ the average cluster size approaches unity, signifying a monomer gas, while
at high $\sigma$ the existence of a liquid phase results in clusters containing essentially all the particles in the simulation.

Kinetic constraints become important at low values of $T$ and $\sigma$. At low $T$ the system becomes unable to escape from incorrect
configurations. Since it very likely that some incorrect bonds will be formed during the assembly process
this results in very low yields, and the system instead
forms glassy kinetic aggregates. At low values of $\sigma$ the patches
are very narrow. Particles will only rarely be sufficiently well aligned to feel attractive interactions from their neighbours, and
growth is suppressed.

\begin{figure}
\includegraphics[width=8.4cm]{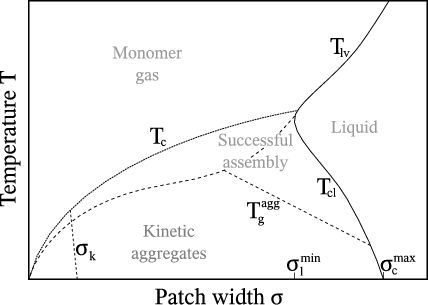}
\caption{\label{fig:SchematicPhases}A schematic diagram illustrating the 
regions of parameter space in which different
behaviours are observed. 
$T_c$ is the midpoint of the transition between the cluster
gas and the monomer gas. 
$T_{lv}$ and $T_{lc}$ are the liquid-vapour phase transition lines for the
monomer and cluster gases, respectively. $T^{\mathrm{agg}}_g$ is the temperature below which aggregates become unable to rearrange
to form target clusters on the time scale of the simulations, instead 
becoming trapped in a glass-like state. $\sigma_k$ is the value of $\sigma$ 
below which the patches are so narrow that little assembly or aggregation is 
able to take place on the time scale of the simulations. 
$\sigma_l^{\mathrm{min}}$ represents the lowest value of $\sigma$ at which the 
liquid state is stable and $\sigma_c^{\mathrm{max}}$ the highest
value of $\sigma$ at which the cluster phase is stable.}
\end{figure}

Sandwiched in the middle of these regions of inhibited assembly is a region of parameter space in which the target is both thermodynamically
stable and kinetically accessible. In this region good yields of the target structure are obtained.  These effects are summarised in Fig.\ \ref{fig:SchematicPhases}, which shows schematically the different regions of parameter space and the
lines which separate them. We will now examine these features in more detail,
focussing on their dependence on the target geometry.

\subsection{The cluster-monomer gas transition}
\label{sec:TcTransition}

As there are only very weak attractive forces between clusters assembled into
the target structures, $T_c$, the temperature above which the stable state is dominated by monomers rather than target clusters, represents the midpoint of
a chemical equilibrium between a monomer gas and a cluster gas.
The plots in Fig.\ \ref{fig:MainHeatPlots} show that $T_c$ increases with patch width $\sigma$ for all of the systems.
This feature can be explained by considering the free energies of the monomer 
and cluster gases.
At $T_c$, $A_m$ and $A_c$, the free energies of the monomer and cluster gases 
respectively, are approximately equal. $A_m$ is largely independent of $\sigma$,
as is the ground state potential energy per particle in the cluster gas, 
$V_c^{gs} \approx m \epsilon/2$,
where $m$ is the number of patches per particle. 
However, the entropy $S_c$ of the cluster gas increases with
$\sigma$. At higher $\sigma$ the clusters are free to undergo larger 
vibrations, leading to a higher vibrational entropy, and hence a larger $T_c$.

\begin{figure}
\includegraphics[width=8.4cm]{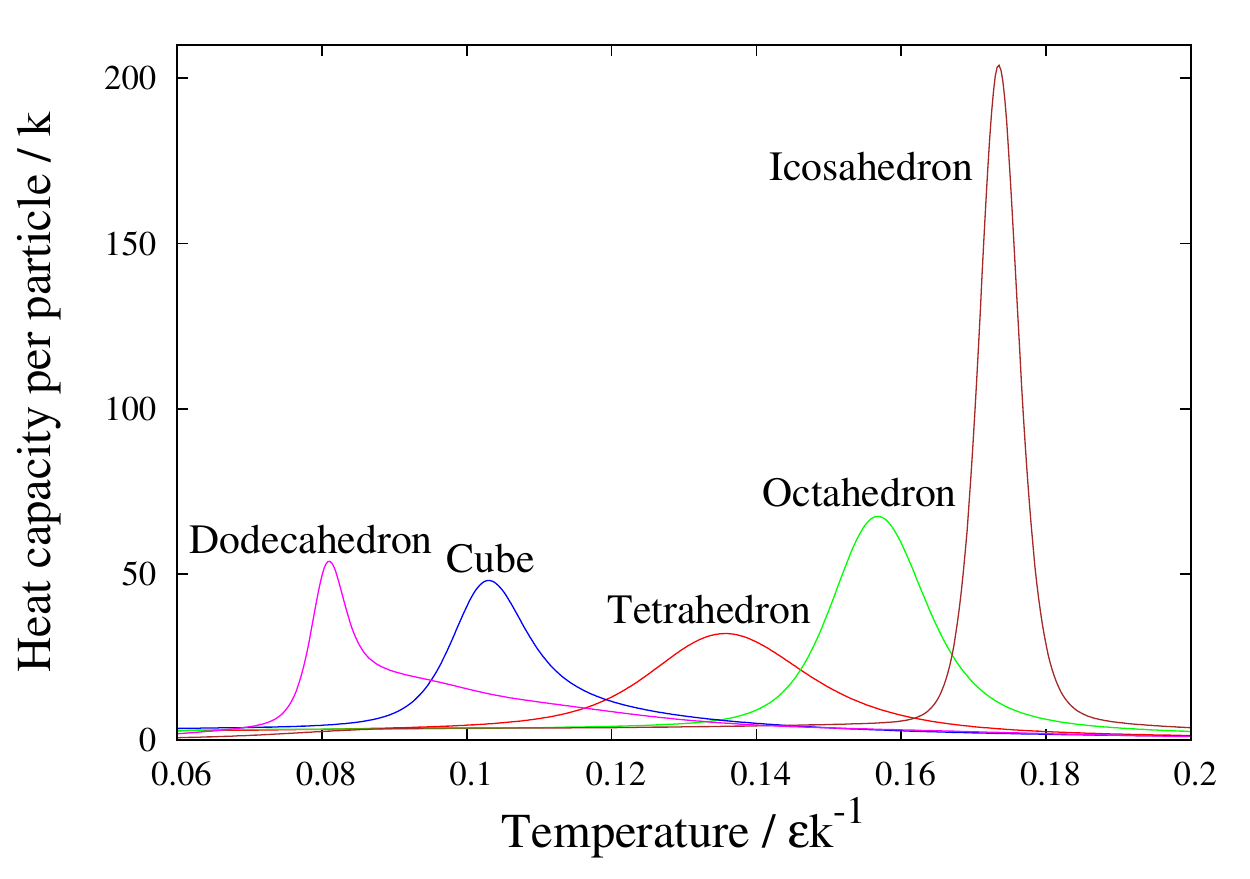}
\caption{\label{fig:AllShapeCvsAnnotated}(Colour Online) The heat capacity $C_v$ as a function of temperature for each of the target structures at
$\sigma = 0.45$.
This data was obtained from umbrella sampling simulations containing sufficient particles to form one target cluster. 
}
\end{figure}

While this trend is a generic feature, $T_c$ also shows strong target 
dependence.  Fig.\ \ref{fig:AllShapeCvsAnnotated} shows the heat capacity
$C_v$ as a function of temperature for each of the target structures 
at $\sigma = 0.45$. The peaks in the $C_v$ curves correspond to the
cluster-monomer transition, and we define $T_c$ as the temperature at 
which $C_v$ is a maximum. 
Note that the $C_v$ plot for dodecahedra exhibits a distinct shoulder to the
right of the heat capacity peak, which is actually indicative
that this system is behaving very differently from the other four. 
We shall examine the dodecahedral system in detail in in 
Section\ \ref{sec:Dodecahedra}.

There are two main target-dependent factors affecting the value of $T_c$. 
Firstly, the number of patches per particle $m$ is important, since the more
patches are present the more the cluster gas is energetically stabilised.
The value of $n$, the number of particles per cluster, has the converse effect. When $N$ particles assemble into $N/n$ clusters there is a reduction in the 
effective number of particles in the system, and hence a corresponding 
reduction in the translational entropy. We can obtain a crude estimate for 
the functional dependence of $T_c$ on $m$ and $n$ if we neglect vibrations 
and ignore the mass dependence of the effective `particles': 
\begin{equation}
T_c 
\propto \frac{m}{1-\frac{1}{n}}.
\label{eqn:TcEstimate}
\end{equation}
Table \ref{table:ShapeCvs} shows estimated values obtained using this equation,
along with values obtained from umbrella sampling simulations.
Despite the simplicity of the model,
it matches the data reasonably well, 
and provides an explanation for the ordering of the $T_c$ values.

\begin{table}
  \begin{center}
    \begin{tabular}{lcccc}
      \hline
      \hline
      & & & \multicolumn{2}{c}{$T_c/\epsilon k^{-1}$} \\
      \cline{4-5}
      Structure & n & m & Measured & Predicted \\ \hline 
      Tetrahedron & 4 & 3 & 0.136 & 0.126 \\ 
      Octahedron & 6 & 4 & 0.157 & 0.151 \\ 
      Cube & 8 & 3 & 0.103 & 0.108 \\ 
      Icosahedron & 12 & 5 & 0.173 & 0.172 \\
      Dodecahedron & 20 & 3 & 0.081 & 0.099 \\ 
      \hline\hline
    \end{tabular}
  \end{center}
  \caption{\label{table:ShapeCvs}Comparison of $T_c$ values at $\sigma=0.45$ obtained 
from umbrella sampling simulations  and estimated using Eq.\ 
\ref{eqn:TcEstimate}. The constant of proportionality in Eq.\ 
\ref{eqn:TcEstimate} was chosen to give the best match to the data. 
}
\end{table}

The value of $T_c$ has an important impact on the ability of a system to
self-assemble, with lower values of $T_c$ making assembly more difficult, 
because the potential range of $T$ over which assembly might occur 
is decreased. It is noteworthy that cubes and dodecahedra have the lowest 
values of $T_c$, and are the hardest of our targets to assemble.

As the monomer-cluster transition is a chemical 
equilibrium, the heat capacity peaks have a finite width. The relative width 
is expected to decrease with increasing target size, and this can be seen
in Fig.\ \ref{fig:AllShapeCvsAnnotated}, 
with the exception of the dodecahedral system, again indicating that this 
system is behaving differently.

\subsection{The cluster-liquid transition}
\label{sec:Aggregates}

For all systems, we see the formation of large aggregates both at
sufficiently large values of $\sigma$ and at low temperature for 
narrower patches (Fig.\ \ref{fig:SizeHeatPlots}). 
The aggregates formed in these two regions
are in fact closely related, with their properties varying smoothly between 
them, as is especially clear for the dodecahedral case,
where there is no region of cluster formation separating the two regions of aggregate formation. 
Nevertheless, it is natural to divide these aggregates into two groups. In the first, at high $\sigma$, the aggregates are the stable state of the system.
In general these thermodynamic aggregates have some liquid-like characteristics, being fairly compact and undergoing constant rearrangement. The second
group of aggregates, those found at low temperature and lower $\sigma$, are metastable, with the target cluster being the true stable state. These kinetic
aggregates also display more gel-like properties, often forming extended networks of string-like formations, which may percolate. However, we have not sought
to establish whether they form as a result of kinetically-arrested liquid-vapour
phase separation, or gelation of a homogeneous fluid.\cite{Sciortino08}

\begin{figure}
\includegraphics[width=7.4cm]{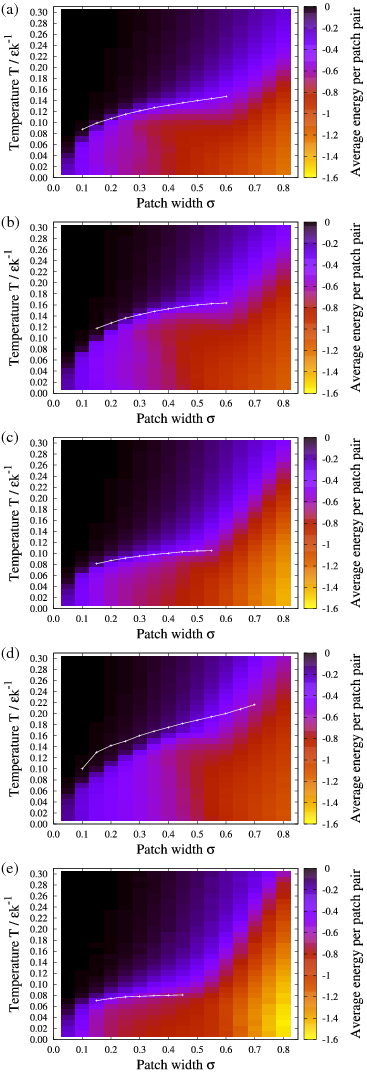}
\caption{\label{fig:EnergyHeatPlots}(Colour Online) The normalised final total energy $E/2Nm$
for systems of particles designed to form (a) tetrahedra, (b) octahedra, 
(c) cubes, (d) icosahedra, and (e) dodecahedra, for the same
simulations as Fig.\ \ref{fig:MainHeatPlots}. The energy is normalised such that a value of $-1$ indicates that, on average, all patches are involved in
one perfect bonding interaction. 
The white lines again show $T_c$.
}
\end{figure}

The aggregates tend to have irregular forms. 
In the case of the thermodynamic aggregates, so long as all patches are 
pointing inwards, the surface tension is relatively low, because the 
energetic cost of the surface is small.
As a result there is a significantly smaller driving force for forming 
spherical droplets than for systems with isotropic interactions. 
In the case of the kinetic aggregates, the reduced ability to rearrange because
of insufficient thermal energy and narrow patches leads to the formation of 
increasingly irregular and ramified aggregates at low $\sigma$. 
The energy of the aggregates tends to become less negative as well, 
since these increasingly glassy systems are less able to optimise their 
configurations to achieve lower energies. This effect is clearly visible in 
Fig.\ \ref{fig:EnergyHeatPlots}, which shows the dependence of the final 
energies in the simulations on $T$ and $\sigma$ for each of the target 
structures.

\begin{figure*}
\includegraphics[width=18cm]{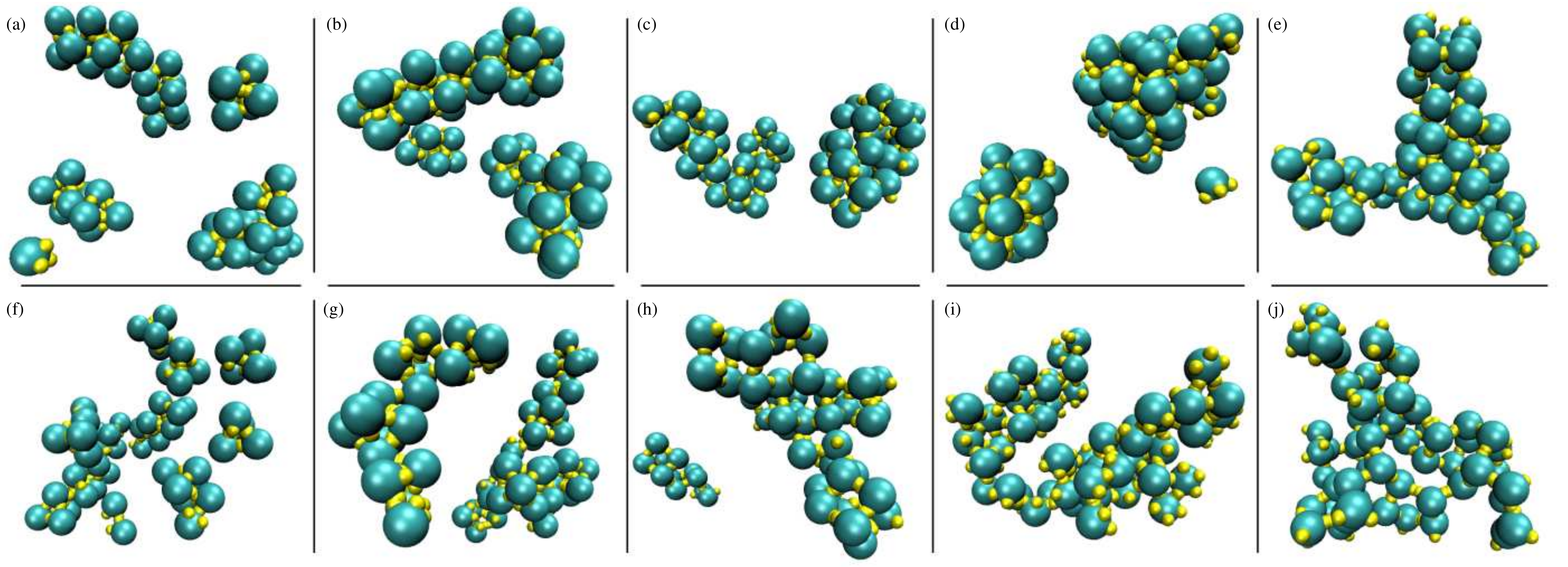}
\caption{\label{fig:AggregatePics}(Colour Online) 
Typical aggregate configurations (a)--(e) in the liquid-vapour coexistence region
 but (a)--(d) close to the region of successful assembly, and (f)--(j) in the kinetic aggregation regime for particles designed to form (a) and (f) tetrahedra, 
(b) and (g) octahedra, (c) and (h) cubes, (d) and (i) icosahedra, and (e) and (j) 
dodecahedra. The values of $(kT/\epsilon,\sigma)$ are (a) $(0.06,0.65)$, 
(b) $(0.09,0.65)$, (c) $(0.08,0.55)$, (d) $(0.16,0.6)$, (e) $(0.04,0.6)$, 
(f) $(0.05,0.35)$, (g) $(0.07,0.35)$, (h) $(0.06,0.4)$, (i) $(0.08,0.35)$ and
(j) $(0.04,0.35)$.
Each simulation contains $60$ particles at a number density of $0.15\,\sigma_{\rm LJ}^{-3}$.
}
\end{figure*}

The structures of the aggregates also have a strong dependence on the target. At temperatures and patch widths close to those where target clusters can
successfully form, the local structure within the aggregates tends to have a 
strong similarity to that in the target cluster, so that icosahedron-like 
groups are visible in the aggregates of icosahedron-forming particles, 
and so on. These structural
similarities are visible in Fig.\ \ref{fig:AggregatePics} (a)--(d), 
which shows snapshots of liquid aggregates near to the region of successful assembly.
The one target which consistently produces aggregates with little resemblance to the target structure is the dodecahedron.
The wide splay angle (i.e.\ the angle between the patches and the symmetry axis) allows the formation of a
wide variety of structures, resulting in highly ramified
structures containing many hexagons and larger rings as well as pentagons (see Fig.\ \ref{fig:AggregatePics}(e)).
However as we move to higher values of $\sigma$ the differences between the aggregates formed for the different targets are lost, so that the dense
aggregates formed at very high $\sigma$ look essentially the same for all cases.

Fig.\ \ref{fig:AggregatePics}(f)--(j) show typical kinetic aggregate structures.
We can see that all of the kinetic aggregates have a tendency to form chain-like structures. In larger
simulations these may sometimes connect together and percolate across the simulation box. The local structure of the chains is again  heavily dependent on
the target structure, reflecting the symmetries of the target. We can also see that the tetrahedra have a greater tendency to form a number of smaller
aggregates, which is consistent with Fig.\ \ref{fig:SizeHeatPlots}. The narrow splay angle of the tetrahedron-forming particles promotes high curvature
and hence small clusters. This effect is lost at higher $\sigma$, as the patch positions begin to have less control over aggregate structure.

Interestingly, similar ramified worm-like aggregates have also been seen for
Janus particles that have a hydrophobic and a charged side.\cite{Hong08}
The similarity is probably because of the effectively one-sided nature of the 
attractions for our patchy particles.

The extent of the regions in which aggregates are found is important, as it sets boundaries on the region in which the target structure may successfully
be formed. In this section we consider the factors affecting the region in 
which liquid-like aggregates are stable, while in Section \ref{sec:Mechanisms}
we will return to consider those determining the extent of the kinetic 
aggregate region.

It is useful to define some terms to describe the position of the transition from the cluster phase to the liquid at high $\sigma$. We define
$\sigma_c^{\mathrm{max}}$ as the maximum value of $\sigma$ at which the cluster phase is found (which will be at zero temperature) and $\sigma_l^{\mathrm{min}}$
as the minimum value of $\sigma$ at which the liquid phase is found. 
These two points are connected by the transition line $T_{lc}$. All three
are shown in Fig.\ \ref{fig:SchematicPhases}. $\sigma_c^{\mathrm{max}}$ depends entirely on the energies of the two states, while the gradient 
of $T_{lc}$ depends on the entropy difference between the two states.

Fig.\ \ref{fig:EnergyHeatPlots} shows that the dodecahedra- and cube-forming particles are the most able to satisfy their patchy interactions in liquid
droplets, and attain the greatest bonding energies. The low number of patches per particle $m$ and the wide angles between patches result in fewer
constraints on the liquid structure, making it easier to optimise patch-patch interactions. We can see a clear
correlation between the angles between patches and the energy of the liquid. For the tetrahedra, octahedra and icosahedra, with angles between the patches of
$60^\circ$, the particles would be required to pack very closely together in order to satisfy all their bonds, and in fact we see relatively small
bonding energies per patch. As we increase the angle, to $90^\circ$ for cubes and then $108^\circ$ for dodecahedra, the energy per patch steadily
increases in magnitude.

The entropy of the aggregates might be expected to be correlated with the energy. When there are fewer constraints it is not only easier to satisfy
patch-patch interactions in a disordered configuration, but there are more ways of doing so, leading to a higher configurational entropy. We would therefore
expect larger aggregate entropies for the dodecahedra- and cube-forming
particles.

We can obtain some information about the relative entropies of the aggregate 
and cluster phases by considering $T_{lc}$. 
The shallower the gradient of $T_{lc}$ the greater the entropy differences 
between the aggregate and cluster phases, with negative gradients indicating 
that the aggregates are entropically favoured. 
Although calculating precise values of $T_{lc}$ would be far from 
straightforward, 
we can get an impression of how $T_{lc}$ varies with $\sigma$ from 
Fig.\ \ref{fig:SizeHeatPlots}. 
From those plots we can see that $T_{lc}$ has a negative gradient for octahedra, icosahedra and probably cubes,
indicating that the liquid has a higher entropy than the cluster gas, and hence that the disorder in the liquid more than compensates for the loss of
translational degrees of freedom on forming a large aggregate. 
$T_{lc}$ is close to vertical for the tetrahedral system indicating  that 
the entropies for the liquid and cluster gas are similar---as the smallest of the targets, the tetrahedral clusters have the largest translational entropy. 
The gradient for the dodecahedral systems is unclear from Fig.\ \ref{fig:SizeHeatPlots}, because of the absence of target formation. 
Later in Sec.\ \ref{sec:Dodecahedra}, we will see that $T_{lc}$ has a far 
shallower slope for dodecahedra, 
which is consistent with our expectations based on the 
low energy of the aggregates and their large cluster size.

\subsection{Mechanisms of assembly and misassembly}
\label{sec:Mechanisms}

\begin{figure}
\includegraphics[width=7.8cm]{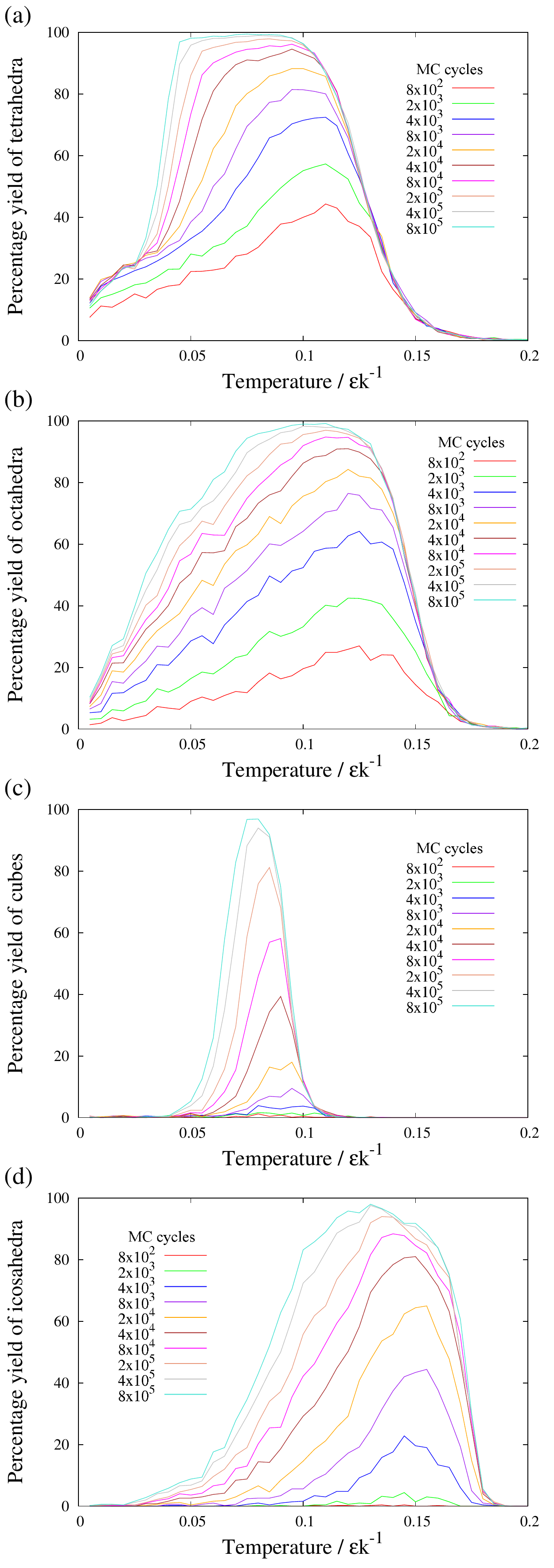}
\caption{\label{fig:AllShapeYieldsWTime}(Colour Online) The yields of
(a) tetrahedra, (b) octahedra, (c) cubes, and (d) icosahedra after different numbers of simulation steps as a function of temperature $T$,
in simulations of systems of 1200 particles at $\sigma=0.45$ and at a number density of $0.15\,\sigma_{\rm LJ}^{-3}$. Each data point is an average
from five simulations.}
\end{figure}

The dynamics of the self-assembly simulations depend strongly on temperature and on the target. Fig.\ \ref{fig:AllShapeYieldsWTime} shows the yields of
each of the target structures as a function of $T$ and time at $\sigma = 0.45$. These plots show a number of interesting features.

Firstly, in each case the yield approaches equilibrium most rapidly at high 
temperatures, close to $T_c$. Here the dynamics are relatively fast, 
and since liquid-like aggregates are not stable with respect to the monomer gas at these temperatures, the clusters face no competition. 
However, close to $T_c$ the yield is limited by the finite width of the 
cluster gas-monomer gas transition, 
with smaller clusters displaying a broader transition, as is clear in 
Fig.\ \ref{fig:AllShapeYieldsWTime}.

At slightly lower temperatures the time taken to reach equilibrium becomes 
longer, both because the time scale for rearrangement and the breaking of
incorrect bonds is longer and because the diffusional time scales increase 
as the assembly progresses towards higher yields. 
As a result the temperature at which optimal yields are obtained decreases 
with time, being determined by the competition between rapid assembly at 
high temperatures and higher equilibrium yields at lower temperatures.

Moving to still lower temperatures the yield on the time scale of our simulations falls off, until around $T^{\mathrm{agg}}_g$ it becomes severely limited.
For target structures of tetrahedra, cubes and icosahedra, reasonably
sharp cut-offs are seen at low $T$, giving approximate values for $T^{\mathrm{agg}}_g$. In the case of the tetrahedra the yield does not fall to zero
but rather to a finite value of around $20\%$, as some yield of tetrahedra is expected by straightforward chance assembly with no rearrangement
required. Unlike the other targets, the plots for octahedra show an approximately linear decay in yields with decreasing temperature. The reasons for this
are unclear.

\begin{figure}
\includegraphics[width=8.4cm]{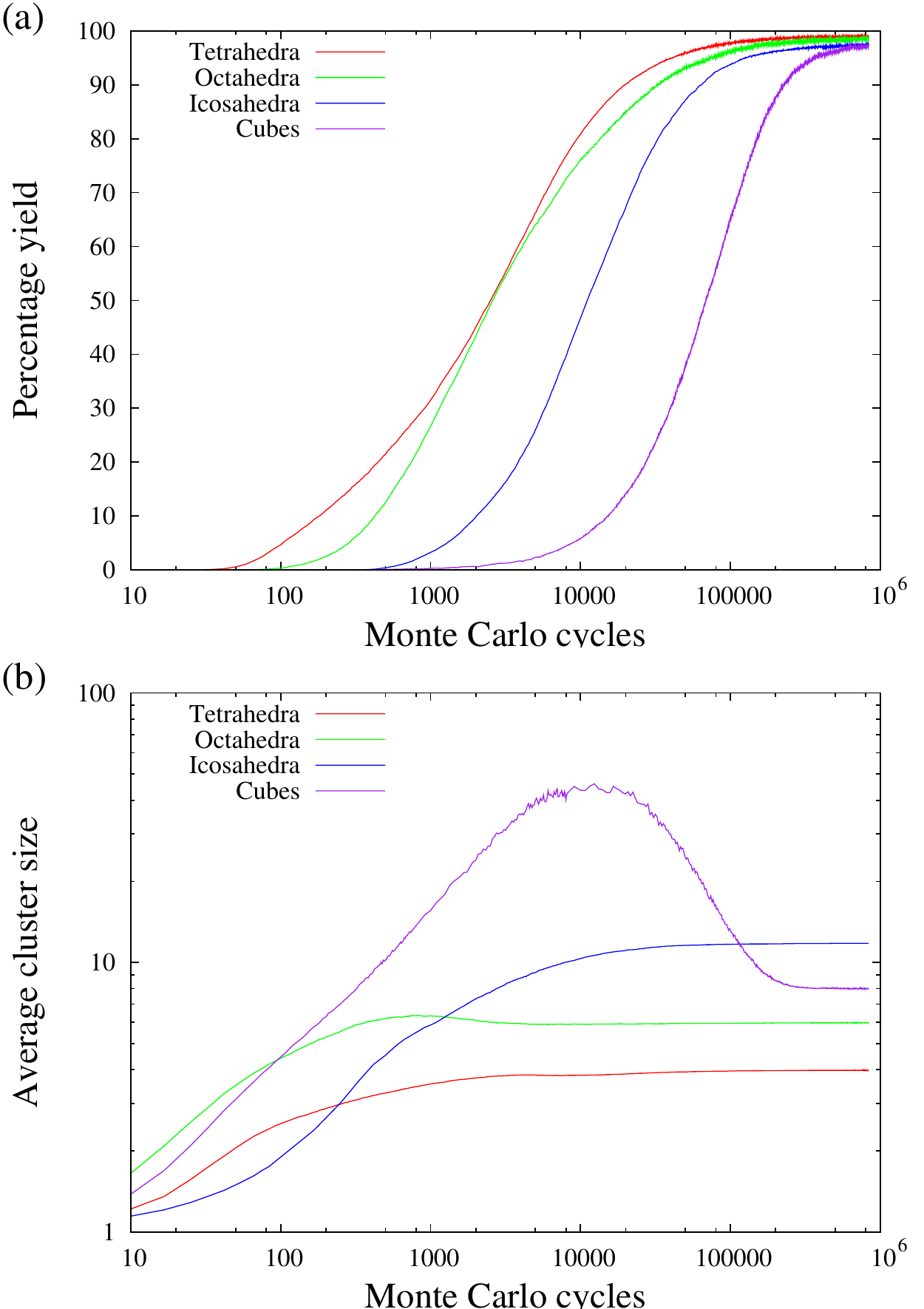}
\caption{\label{fig:Optimal}(Colour Online) (a) Yields and (b) average 
cluster sizes (weighted over particles) as a function of time at 
the optimal conditions for assembly of tetrahedra, octahedra, icosahedra and 
cubes, averaged over 100 simulations. The parameters used were 
$T=0.09\epsilon k^{-1}$, $\sigma=0.4$ for tetrahedra, 
$T=0.11\epsilon k^{-1}$, $\sigma=0.5$ for octahedra, 
$T=0.14\epsilon k^{-1}$, $\sigma=0.45$ for icosahedra and 
$T=0.08\epsilon k^{-1}$, $\sigma=0.45$ for cubes.}
\end{figure}

Fig.\ \ref{fig:Optimal}(a) shows the yield as a function of time for each target structure (except the dodecahedron) at the optimal conditions for the 
formation of that structure in Fig.\ \ref{fig:MainHeatPlots}
(i.e.\ those that give the maximum yield after $80\,000$ cycles).
While the final yields obtained in each case approach $100\%$, the time taken varies by almost two orders of magnitude. The order of the time taken for
tetrahedra, octahedra and icosahedra is consistent with their relative sizes, with larger clusters taking longer to assemble as expected. Some tetrahedra
form at very short times via the chance assembly mentioned above. However, the cubes take anomalously long to assemble, which is partly explained by the 
narrow range of temperature over which cubes can assemble 
(Fig.\ \ref{fig:AllShapeYieldsWTime}(c)) because of the low value of $T_c$ for 
this system. 

We  will now consider in more detail the dynamics in 
the region of parameter space where the target clusters are most stable
by considering three sub-regions of that space. 
Firstly, there is the regime where the target cluster is the only structure 
stable with respect to the monomer gas (i.e.\ $T_{lv} < T < T_c$). 
As a result assembly occurs by direct nucleation of the target structure by the
addition of monomers and other small clusters to a growing cluster, 
rather than proceeding via an aggregate state. 
Because aggregates are not stable there are essentially no competing states to
the target structure.

Note that as this region is close to $T_c$, the yield may be constrained by the finite width of the cluster-monomer transition, i.e.\ the equilibrium
yield of clusters may be significantly less than $100\%$. Lower temperatures will give a higher equilibrium yield.
In this region the speed of assembly 
becomes slower with decreasing $\sigma$ due to the comparative rarity of 
bonding events between particles with narrow patches.

In the second regime both the target clusters and large liquid aggregates are stable with respect to the monomer gas (i.e.\ $T < T_c$ and $T < T_{lv}$ but
$T>T^{\mathrm{agg}}_g$), leading to competition. 
However, mobility in the liquid droplets allows them to rearrange to form 
target clusters. Once complete clusters are formed all their patches point 
inwards, such that they experience very little attractive interaction with 
other particles. They are therefore able to ``bud off'', separating from the 
rest of the liquid droplet in which they formed.

The optimal conditions for assembly are typically found close to the boundary of
these two regimes, at a point in parameter space where both the direct 
nucleation and ``budding-off'' mechanisms of assembly operate
Fig.\ \ref{fig:Optimal}(b) shows the average cluster size against time under 
optimal assembly conditions for each of the target structures. 
For the cubes (and to some extent the octahedra) the average size passes 
through a maximum before falling to the target cluster size, indicating clearly 
that the mechanism of aggregation followed by rearrangement plays a significant
role. 
Even though the ``budding off'' mechanism is not sufficiently prominent to cause
a maximum in the cluster size for tetrahedra and icosahedra, it still 
contributes to the rapidity of assembly under these conditions.

For $T<T^{\mathrm{agg}}_g$ assembly is suppressed because the aggregates act
as kinetic traps. Hence, we define $T^{\mathrm{agg}}_g$ as the temperature 
below which large clusters have insufficient thermal energy to rearrange to 
form target clusters on the time scale of the simulation
(and is therefore weakly dependent on time scale). 
Below $T^{\mathrm{agg}}_g$ generally only large aggregates are formed, 
often forming extended ramified networks which may percolate.

$T^{\mathrm{agg}}_g$ is dependent on target geometry for a number of reasons. 
The first factor is the degree of order in the aggregates.
Fig.\ \ref{fig:AggregatePics} shows
that the aggregates can contain a considerable degree of local structure
that is similar to the target, and hence only a relatively small amount of 
rearrangement is required for the aggregates to form complete clusters. 
The exception is the dodecahedral case, 
where the aggregates have very little dodecahedron-like structure at all, 
and would require almost total rearrangement to form the target clusters. 

A second factor is the stability of the aggregates, because the more stable 
they are, the smaller is the thermodynamic driving force for them to rearrange
to find the target structure. This factor is particularly relevant to the 
cubic and dodecahedral systems, which we saw earlier, had particular low
energies and high entropies.

A third factor is the size of the aggregates. 
In general a smaller splay angle in the assembling particles
tends to lead to a higher surface curvature and hence to smaller aggregates,
which are easier to rearrange. In Fig.\ \ref{fig:SizeHeatPlots} we can see that 
the low-$T$ aggregates formed with tetrahedra and octahedra as 
target structures are much smaller than for the other shapes.

A final factor is the size of the target structures themselves. 
Less rearrangement is needed to form a smaller target. 
Further, for smaller targets, the growth of aggregates is less likely to 
deviate significantly from the correct assembly pathway. 
For example, for tetrahedra, some degree of successful
assembly is observed even at very low $T$, because a significant proportion 
tetramers will have come together by chance to directly form a tetrahedron 
without the need for any rearrangement. However, even for octahedra this 
effect becomes negligible.

\subsection{The difficulty of assembling dodecahedra}
\label{sec:Dodecahedra}

Of the target structures examined in this paper, the dodecahedron is exceptional in the difficulty of its assembly. While the other challenging target, 
the cube, shows a restricted region in which assembly occurs, it is nevertheless easy to obtain high yields by conducting long simulations with carefully
chosen parameters. By contrast dodecahedra are never formed with our model potential under any set of conditions.

In order to more fully understand the difficulty of assembling dodecahedra we attempted to obtain equilibrium thermodynamic data for a system of
20 particles, sufficient to assemble one dodecahedron. This was achieved only with considerable difficulty. Indeed, the pathological difficulty
of forming a single dodecahedron even using biasing techniques underlines the extreme unlikeliness of ever forming one under ordinary dynamics, and
implies the presence of a large kinetic barrier to assembly.
Success was eventually achieved by combining two-dimensional umbrella sampling (using the number of bonds and the number of pentagons in the largest
cluster as order parameters) with Hamiltonian exchange
as described in Section \ref{sec:EquilSims}. 

In general, because of the low energy of the target structures and the
specificity of the interactions, one would imagine that at moderate $\sigma$ 
there must exist a temperature range over which the target is the only stable 
structure with regard to the monomer gas, and so where aggregate 
formation does not compete with the assembly of the target.
One might expect that in this region at least, dodecahedra should be able to 
assemble, so long as the time scales are sufficiently long that the nucleation
barrier can be overcome. However, as we will see, this is in fact not the case, 
since over most of the range of $\sigma$ this region does not exist for the 
dodecahedral system. 

\begin{figure}
\includegraphics[width=8.4cm]{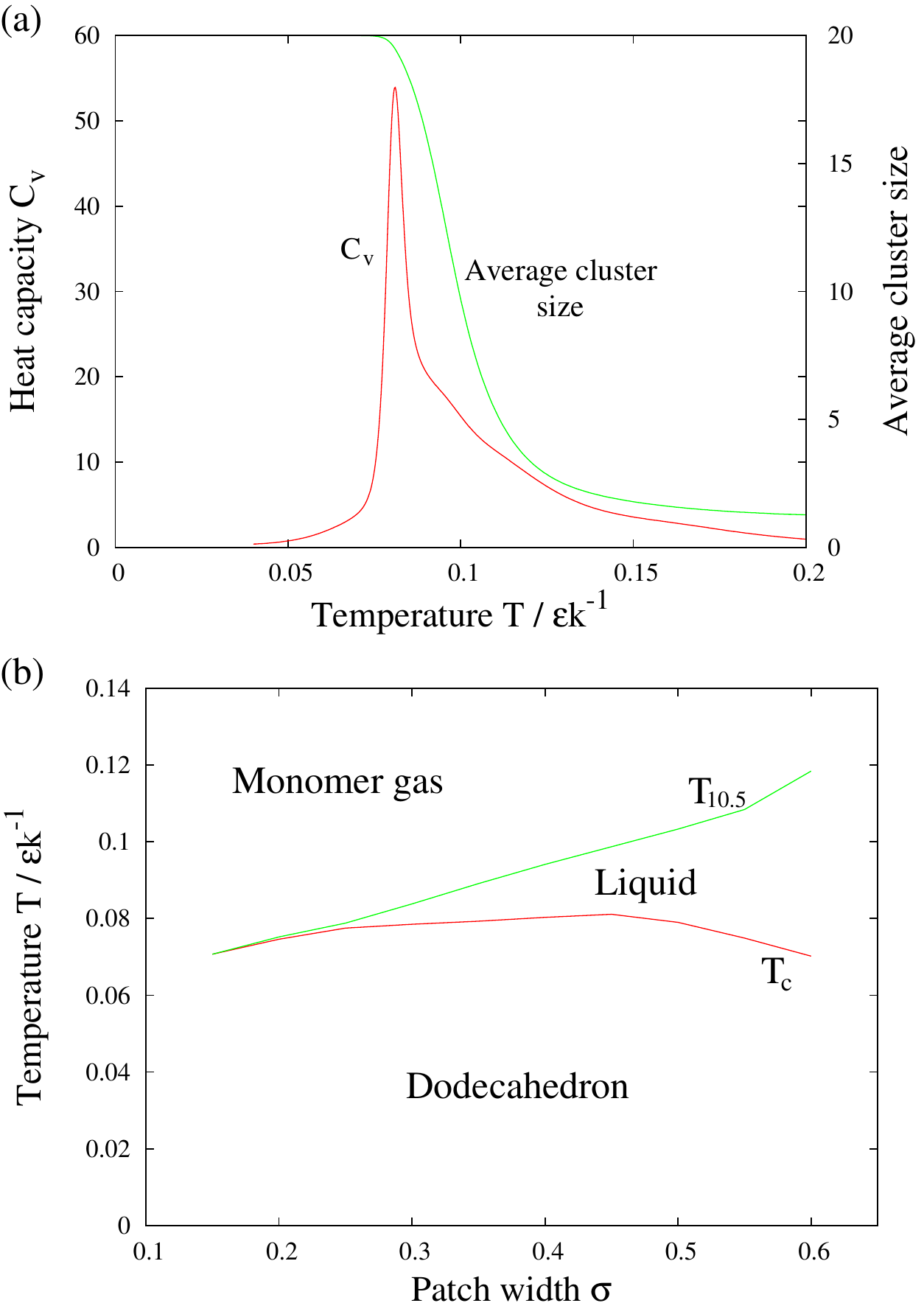}
\caption{\label{fig:DodecaEquil}(Colour Online) 
Equilibrium properties of a system of 20 dodecahedron-forming particles 
in the canonical ensemble. 
(a) The heat capacity $C_v$ and average cluster size (weighted by particles) 
as a function of temperature. 
(b) $T_m$ (the temperature at the peak in $C_v$) and $T_{10.5}$, 
(the temperature at which the average cluster size is 10.5) as a function of 
$\sigma$.
}
\end{figure}

Fig.\ \ref{fig:DodecaEquil}(a) shows a plot of the heat capacity $C_v$ along 
with a plot of the average cluster size for $\sigma = 0.45$. At temperatures
below the peak in $C_v$ the dodecahedron is most stable. 
However, we can see that the average cluster size remains high beyond this
point, indicating that the dodecahedron first melts before gradually
evaporating at higher temperature. The evaporation of this liquid cluster
corresponds to the shoulder in the $C_v$ plot.

The generality of this picture is confirmed in Fig.\ 
\ref{fig:DodecaEquil}(b). 
The higher line in the figure shows the temperature at which the average cluster size in the box has a value of $10.5$ (the midpoint of $1$ and $20$), 
and the lower line $T_m$, the temperature at which a dodecahedron melts. 
The presence of an intermediate liquid state for this 20-particle system,
 which persists down to $\sigma\approx 0.25$, was a surprise to us. For, 
although we expected the free energies of the monomeric and cluster states to 
be relatively little affected by the small number of particles in these 
equilibrium simulations, we expected the liquid state to be considerably 
destabilized by the high surface to volume ratio of any 20-particle liquid 
cluster. That liquid clusters are seen in these simulations is testament
to the low energy and high entropy of the aggregates for the dodecahedral
system that we noted earlier. Thus, we expect the region of stability for the
liquid state to be larger for a bulk system than for this 20-particle system, 
i.e.\ $T_{10.5}$ can be considered as a lower bound to $T_{lv}$, $T_m$ as 
an upper bound to $T_{lc}$ and 0.25 is an upper bound for $\sigma_l^{\rm min}$

These results thus allow us to begin to understand why the dodecahedra are so
hard to assemble.
Given that at moderate $\sigma$ the stable state of the system at higher temperatures is as liquid droplets, the only region in which dodecahedra might be
able to form is at lower temperatures, where, although they represent the 
global free energy minimum, aggregate formation competes with target assembly. 
In particular, in this region the system will aggregate very quickly, 
and so for a dodecahedron to form these aggregates would then need to be
be able to rearrange sufficiently, a process which is severely inhibited by 
the slow dynamics at these low temperatures. 
For the dodecahedral system though, such rearrangement appears to be far more 
difficult than would be expected even allowing for temperature. This is probably
a result of geometric factors which are unique, within our set of targets, to the dodecahedron. Because of the widely spaced patches on the particle
surfaces, an enormous range of incorrect configurations are possible which satisfy the bonding well, and which deviate significantly from the target
structure even on a local scale. For example, as well as pentagons, (non-planar) hexagons or larger polygons can form without introducing strain. Furthermore,
it is possible for bonds to pass through these rings, leading to entangled 
networks of particles.
These structures have little in common with correctly 
formed dodecahedra, and further, they experience little pressure to
rearrange.

\begin{figure}
\includegraphics[width=8.4cm]{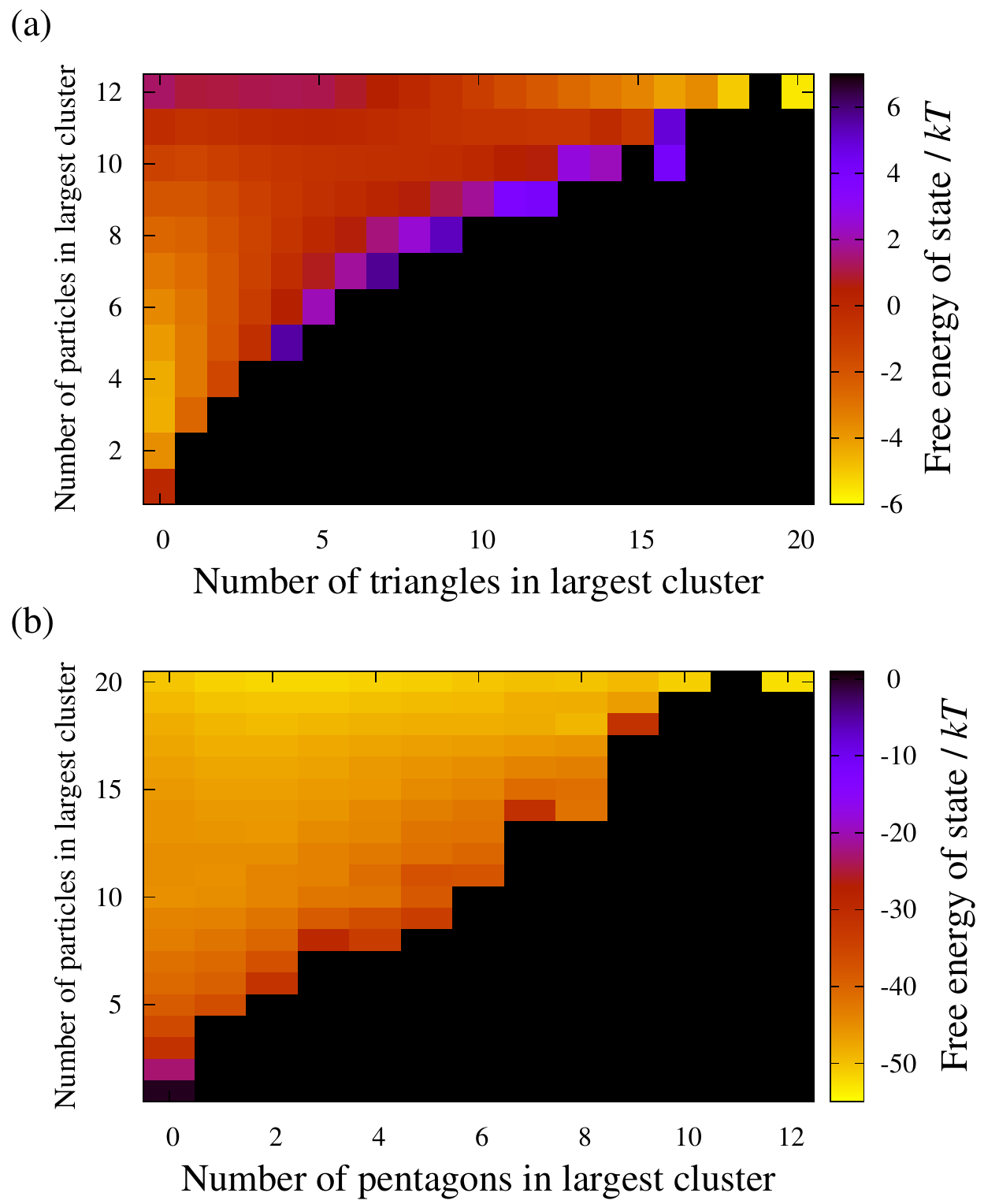}
\caption{\label{fig:DLandscapes}(Colour Online) Free energy landscapes 
for systems of (a) 12 icosahedron-forming particles, 
and (b) 20 dodecahedron-forming particles at $T=T_c$ 
and $\sigma=0.45$. The order parameters are the numbers 
of particles and triangles/pentagons in the largest cluster. 
}
\end{figure}

\begin{figure}
\includegraphics[width=8.4cm]{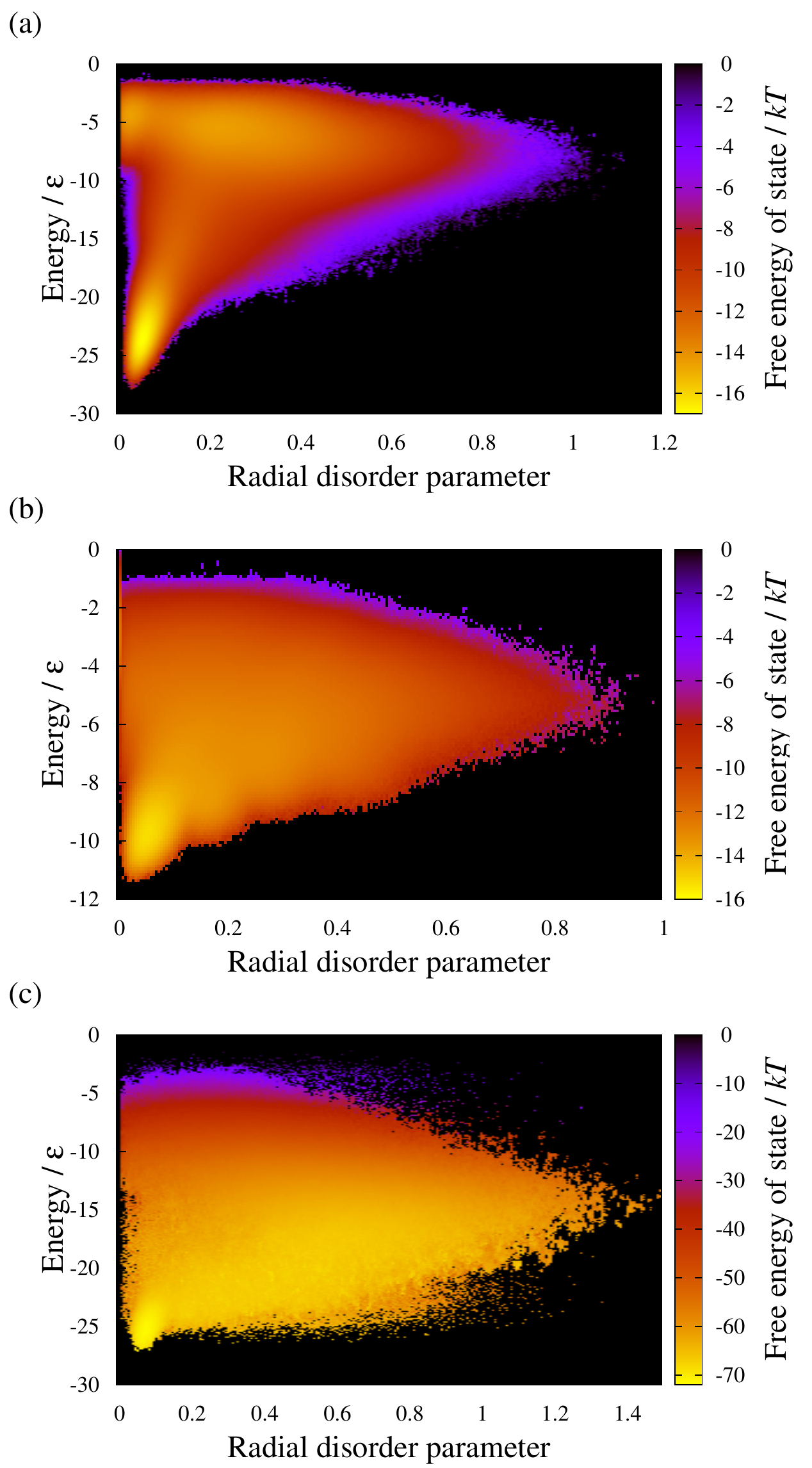}
\caption{\label{fig:CLandscapes}(Colour Online) Free energy landscapes 
for systems of (a) 12 icosahedron-forming particles, (b) 8 cube-forming particles
and (c) 20 dodecahedron-forming particles at $T=T_c$ 
and $\sigma=0.45$. The order parameters are
the configurational energy and a measure of radial disorder, described in the text. 
}
\end{figure}

In order to obtain a clearer picture of the thermodynamic constraints on the 
dynamics of the system, we made use of data from the Hamiltonian exchange 
simulations to plot free energy landscapes in Fig.\ \ref{fig:DLandscapes} and \ref{fig:CLandscapes}, not only for dodecahedra but also for icosahedra and cubes.
The icosahedral system provides a contrasting example of a successfully assembling 
system.  All are at a temperature corresponding to the maximum in the heat capacity. 


Fig.\ \ref{fig:DLandscapes} shows the free energy as a function of the number of 
particles and the number of correct polygons (triangles for the icosahedron and 
pentagons for the dodecahedron) in the largest cluster in the system. 
In Fig.\ \ref{fig:DLandscapes}(a)
we can see clear free energy minima representing the monomer gas and 
the complete icosahedron. 
Between the two minima is a region of higher free energy, with a 
free energy transition state with a significant amount of triangular ordering. 
Once the transition state is passed, the shape of the landscape directs the 
system towards forming an icosahedron, facilitating assembly. 
(Note that the high free energy of clusters of 12 particles with 19 triangles
simply represents the fact that if one bond of an icosahedron is broken two 
triangles are lost, and does not represent a barrier.)

By contrast, the monomeric gas is not a free energy minimum for the dodecahedral
system at temperatures at which the dodecahedron is most stable (Fig.\ 
\ref{fig:DLandscapes}(b)). There is no barrier to the formation
of a 20-particle cluster, so the system will tend to rapidly aggregate, and 
will then experience little drive to form a dodecahedron. Indeed, if the system
follows the steepest downhill path it will tend to form a single aggregate containing few pentagons, so that significant rearrangement would be needed
to form a dodecahedron.

Fig.\ \ref{fig:CLandscapes} shows the free energy as a function of two different order parameters, the potential energy and a radial disorder parameter. 
The latter is defined as the standard deviation in the distance of particles from the centre of mass, and
has a value of zero when all particles lie on a spherical shell (such as in the 
target structures). Again, the plot for the icosahedron shows two
free energy minima, corresponding to a monomer gas and to an icosahedron, 
separated by a transition state of higher free energy. Importantly, we see 
that in order to obtain lower energies the system is forced to steadily reduce 
its maximum radial disorder, such that it is directed into the icosahedral 
state. The energy landscape is funnel-like.\cite{Bryngelson95}

By contrast, the free energy of the dodecahedral system appears to be almost entirely a function of the configurational energy, with very little
dependence on the radial disorder. Energies close to that of the global minimum can be reached by highly disordered states. As a result the free energy 
landscape is relatively flat, and the system experiences very little driving 
force to form an ordered cluster. 
We also depict the landscape for cube-forming particles in Fig.\ \ref{fig:CLandscapes}(b). The intermediate difficulty of assembling the cubes is reflected in the 
broader, less-funneled nature of the lanscape compared to the icosahedron, but
there is still some free energy gradient towards the target structure. 

Nevertheless, neither of the dodecahedral plots fully explain the situation. From these diagrams one might expect the formation of dodecahedra to be rare,
but not essentially impossible. It appears that there is a large kinetic barrier preventing the formation of dodecahedra, which is not visible in these
diagrams due to the choice of order parameters. 
Despite considerable efforts, we were not able to find a combination of 
order parameters which would show this barrier explicitly.

\section{Conclusions}

We have presented a study of the self-assembly of a class of simple targets, the Platonic solids, using a minimal model of assembling patchy particles. We
have comprehensively mapped out the behaviours of systems of particles designed to form each of the targets as a function of temperature and patch width.
Further, we have obtained equilibrium data using umbrella sampling and Hamiltonian exchange, allowing us to compare the thermodynamic properties of the
different systems, and to produce free energy landscapes for the larger targets to elucidate the differences in their behaviour.

We find that the behaviour of the systems and the success of self-assembly are strongly dependent on the target structure. One key property which
varies considerably between targets is the stability of disordered aggregates, which have a complex relationship with the self-assembly
process. Over large regions of parameter space they act as competition and prevent successful assembly, but in other regions assembly may proceed
by first forming aggregates which then rearrange to ``bud off'' the target structures. As a result the assembly of cubes, whose constituent particles form
particularly stable aggregates, is dominated by this budding mechanism, while dodecahedron-forming particles form aggregates so stable that they
effectively block assembly of dodecahedra. In general we find that the stability of aggregates relative to the target clusters is determined
largely by the spacing between the patches on the particles' surfaces. Wider patch spacing leads to fewer constraints on the structures that the aggregates
can adopt while satisfying all the patchy interactions, which in turn leads to both lower energies and higher entropies.

The stability of the target clusters themselves also varies considerably between targets, and depends both on the number of interactions each particle
participates in, and on the size of the cluster, where greater size tends in general to decrease stability. 
High stability is a desirable trait as it allows assembly at high temperatures
(or equivalently for weak interactions), where the breakup and rearrangement of misformed structures is more rapid.

We find that the dependence on shape can be such that certain shapes never assemble successfully; dodecahedra appear to be essentially impossible to
correctly assemble in our model. The dodecahedron is a relatively unstable target because of its small number of interactions (each particle having only
three nearest neighbours) and large size. The aggregates with which it competes, by contrast, are exceptionally stable.
As a result at moderate patch widths the aggregates are stable to higher temperatures than the dodecahedra, and indeed there is no region of
parameter space where the dodecahedron is both thermodynamically stable and kinetically accessible.

These observations collectively suggest a number of design rules for targets that will assemble easily. Firstly, a high number of nearest neighbours
allows for more interaction which stabilise the target structure. Secondly, closely spaced patches serve to inhibit the formation of stable
aggregates which can provide competition. These two features can be combined by choosing shapes with triangular faces, and as such we would argue that
triangular faces are a major advantage in any target structure.

Proceeding to larger and more ambitious self-assembly targets, a difficulty arises. As the target size increases, the competition between the target
and disordered aggregates increasingly favours aggregates, as the targets become less entropically favourable. This suggests limits to the size
and complexity of targets of monodisperse self-assembly which can be successfully formed using the simple interactions we have used here. 
One approach to circumvent this difficulty might be to use multi-component systems 
with different patch types that interact selectively, the latter being potentially
achievable using DNA-mediated interactions between the particles.\cite{Maye09}
An alternative approach to this difficulty would be to introduce further constraints to the potential. In particular a potential including torsional constraints, i.e.\
one in which the interactions are specific both in direction and in relative orientation, would massively reduce the number of competing configurations available.
Protein-protein interactions have this kind of specificity, and we expect that this is crucial in the assembly of large structures such as virus capsids.
We examine the effect of using such a modified potential in the accompanying paper.\cite{accompanying}

\begin{acknowledgments}
The authors are grateful for financial support from the EPSRC and the Royal Society.
\end{acknowledgments}

\end{document}